\documentclass[aip,amsmath,amssymb,reprint]{revtex4-1}

\usepackage{graphicx}
\usepackage{grffile}
\usepackage{xcolor}
\usepackage{dcolumn}
\usepackage{bm}

\usepackage[utf8]{inputenc}
\usepackage[T1]{fontenc}
\usepackage{mathptmx}
\usepackage{etoolbox}
\usepackage{hyperref}
\usepackage{overpic}

\renewcommand{\d}{{\rm d}}
\newcommand{\half}{{\frac{1}{2}}}

\newcommand{\rev}[1]{\textcolor{red}{#1}} 
\renewcommand{\rev}[1]{#1} 
\newcommand{\ignore}[1]{}

\newcommand{\avg}[1]{\left\langle{#1}\right\rangle}

\newcommand{\alphaSN}{{\alpha_\text{\,SN}}}
\newcommand{\alphaTC}{{\alpha_\text{\,TC}}}

\newcommand{\N}{{\mathbb{N}}}
\newcommand{\Z}{{\mathbb{Z}}}

\makeatletter
\def\@email#1#2{%
 \endgroup
 \patchcmd{\titleblock@produce}
  {\frontmatter@RRAPformat}
  {\frontmatter@RRAPformat{\produce@RRAP{*#1\href{mailto:#2}{#2}}}\frontmatter@RRAPformat}
  {}{}
}%
\makeatother
\begin{document}

\preprint{AIP/123-QED}


\title{\rev{Non-local transitions and ground state switching in the self-organization of vascular networks}}

\author{Konstantin Klemm}
\email{kklemm@posteo.net}
\affiliation{Instituto de F\'isica Interdisciplinar y Sistemas Complejos (IFISC, CSIC-UIB), Campus Universitat de les Illes Balears,
E-07122 Palma de Mallorca, Spain}

\author{Erik A. Martens}%
\email{erik.martens@math.lth.se}
\affiliation{ Centre for Mathematical Science, Lund University, Sölvegatan 18B, 22100, Lund, Sweden}%

\date{\today}

\begin{abstract}
The model by Hu and Cai [Phys. Rev. Lett., Vol. 111(13) (2013)~\cite{HuCai2013}] describes the self-organization of vascular networks for transport of fluids from source to sinks. Diameters, and thereby conductances, of vessel segments evolve so as to minimize a cost functional $E$. The cost is the trade-off between the power required for pumping the fluid and the energy consumption for vessel maintenance. The model has been used to show emergence of cyclic structures in the presence of locally fluctuating demand, i.e. non-constant net flow at sink nodes. Under rapid and sufficiently large fluctuations, the dynamics exhibits bistability of tree-like and cyclic network structures. We compare these solutions in terms of the cost functional $E$. Close to the saddle-node bifurcation giving rise to the cyclic solutions, we find a parameter regime where the tree-like solution rather than the cyclic solution is cost-optimal. \rev{Thus we discover an additional, non-local transition where tree-like and cyclic solutions exchange their roles as minimum-cost (or ground) states.} The findings hold both in a small system of one source and a few sinks and in an empirical vascular network with hundreds of sinks. In the small system, we further analyze the case of slower fluctuations, i.e., on the same time scale as network adaptation. We find that the noisy dynamics settles around the cyclic structures even when these structures are not cost-optimal.
\end{abstract}

\maketitle

\begin{quotation}
Transport or distribution networks are crucial for the proper functioning of a wide range natural and technological systems. An important question is what network structures are optimal in terms of energy consumption. A classical result is that optimal transport networks are minimal spanning trees, provided that loads are stationary~\cite{Kantorovich1942translocation}; however, many realistic settings display fluctuating loads, in which case cyclic motifs allow for short cuts (shunts), thus improving energy consumption. A number of studies demonstrated the necessity of such cycles using optimization techniques~\cite{Katifori2010,Dodds2010,Corson2010,Bohn2007,Durand2007}; an alternative dynamical systems approach was put forward by Hu and Cai~\cite{HuCai2013} in a model where vessel diameters adapt dynamically in response to a power dissipation function, leading to adaptive network dynamics~\cite{berner2023adaptive}. The bifurcation structure for this model has been subject of a number of studies~\cite{HuCai2013,MartensKlemm2017,martens2019cyclic,KlemmMartens2023}, which assumed that fluctuations occur very rapidly, i.e., time scales associated with fluctuations and the vessel conductance were perfectly separated. Here, we relax this assumption and use energy methods to elucidate the role of slower fluctuations and show that in parameter regimes, where bi-stable configurations with tree-like and cyclic motifs, noise may render cyclic motifs even though tree-like configurations are energetically favorable.
\end{quotation}


\section{Introduction}

In natural and engineered systems, transport and flow networks are essential to the distribution of resources~\cite{banavar1999size,rohden2012self}. Transport networks range from mammalian vascular vessel networks~\cite{Blinder2013,kirst2020mapping}, leaf venation~\cite{ronellenfitsch2015topological}, and brain glymphatic fluid pathways~\cite{kelley2022glymphatic} to arboreal sap conduits~\cite{jensen2016sap}; and man-made infrastructures like roads~\cite{lammer2006scaling} and microfluidic circuits~\cite{abgrall2007lab}. Transport networks are tasked with the critical function of efficiently channeling resources from sources to sinks amidst diverse demands and constraints. The efficiency of these networks is often attributed to their ability to self-organize and adapt dynamically to local flow conditions~\cite{Gross2008,berner2023adaptive}, resulting in complex network topologies that incorporate cycles and loops~\cite{Katifori2010,Dodds2010,konkol2022interplay,kaiser2020discontinuous} and community structures~\cite{kaiser2022dual}. Such configurations are not only indicative of robustness and flexibility, but are also optimized in terms of energy consumption.

While the topology of technological transport networks tends to be more static and thus offer limited adaptability, biological networks such as the mammalian vascular system demonstrate remarkable flexibility and operate reliably over a vast parameter range~\cite{gibbons1994emerging,jacobsen2009tissue,Postnov2016}, thereby avoiding the risk of failure even in extreme scenarios. In the mammalian vasculature, vessels dynamically adjust diameters in response to changes in flow properties like pressure and shear stress~\cite{Jacobsen2003,jensen2013vascular}. 

A range of studies have been dedicated to the optimal network configuration in terms of energy consumption. While an optimal  distribution network assumes the structure of a minimal spanning tree when loads are static~\cite{Kantorovich1942translocation}, a range of studies revealed that fluctuating loads instead result in the emergence of cyclic motifs in order to optimize energy consumption~\cite{Katifori2010,Dodds2010,Corson2010,Bohn2007,Durand2007}. Intuitively, this can be explained by the realization that shunting (shortcutting) nearby nodes is cheaper than rerouting the fluctuating demands up and down a number of levels in a tree structure. An alternative line of research, introduced by Hu and Cai~\cite{HuCai2013} takes the perspective of dynamical systems theory, instead considering the time evolution of the network by invoking a dynamic rule for vessel diameters (conductances), such that they adapt in response to a power consumption function controlled by an exponent $\gamma$. Bifurcation analysis for a minimal triangular motif and larger networks in the limit of rapid fluctuations revealed that cyclic motifs may emerge in either transcritical and saddle-node bifurcations, depending on the specific value of $\gamma$; moreover, the saddle-node bifurcations give rise to bi-stability between tree-like and cyclic configurations~\cite{MartensKlemm2017,martens2019cyclic,KlemmMartens2023}.
In such a dynamical formulation, the time scales of the adaptation rule for the conductances and of fluctuations are of order $\sim 1$ and $\sim T$, respectively --- thus, a time scale separation is made explicit. Crucially, biological settings do not adhere to a \emph{strict} time scale separation, i.e., we cannot claim $T\ll 1$. This raises the question how the time scale associated with fluctuations, $T$, affects bifurcations and bi-stability. Specifically, previous studies used the model assumption of rapid fluctuations, $T\ll 1$, but it is unclear to what extent this assumption accurately predicts the favored network configurations.  

This article seeks to understand the response of adaptive flow networks, subject to non-rapid fluctuation. To achieve this, we first derive a framework for the rapid fluctuation limit, using an 'energy function'~\cite{strogatz2018nonlinear,HuCai2013},  in physical terms describing the energy consumption due to dissipation along the total network. A minimal energy principle~\cite{Lu2022} is used to characterize the stability of equilibria and offers a more global view on stability. Using this methodology, we are then able to study how the timescale and level of fluctuations affect the system dynamics under various conditions. Our study reveals that fluctuations may stabilize cyclic configurations even when tree-like configurations are energetically favorable. Thus, we clarify the influence of the time scale associated with fluctuations, an aspect which was neglected in previous studies.


\section{Model}

\subsection{The model by Hu and Cai}

Let $V$ denote a set of nodes of a network with $N = |V| < \infty$ and $A \subseteq N \times N$ the set of edges. The edges are bidirectional; therefore, $(i, j) \in A$ implies $(j, i) \in A$. Each node is assigned pressure $p_i$. The edge flow is $Q_{ij} > 0$ from node $i$ to $j$. We assume that the network is resistive and linear, i.e., Ohmian with $Q_{ij} = C_{ij}(p_i - p_j)$, where an edge carries the property of conductance between nodes $i$ and $j$ with $C_{ij} = C_{ji} > 0$ only if $(i,j)\in A$. 

At each node $i$, Kirchhoff's law (mass balance) demands that
\begin{align}\label{eq:massbalance}
h_i = \sum_{j \in V} Q_{ij} = \sum_{j \in V} C_{ij} (p_i - p_j)
\end{align}
with 
the constraint  guaranteeing that total mass be conserved, 
$\sum_{i \in V} h_i =0$.

Given conductances $C_{ij}$ and flows $Q_{ij}$, the energy consumption of the system is postulated as
\begin{align}\label{eq:Energy_general}
 E &= \half \sum_{(k,l)\in A} \frac{Q^2_{kl}}{C_{kl}}+c_0 C_{kl}^\gamma
\end{align}
by Hu and Cai~\cite{HuCai2013}. Each vessel segment, represented by an edge $(k,l)$, consumes energy in terms of two amounts in Eq.~\eqref{eq:Energy_general}. The first contribution is the power required to sustain the flow $Q_{kl}$, akin to the power $R I^2$ dissipated by an electric current $I$ over a resistance $R= 1 /C$. Secondly, the power required for maintenance of the vessel segment depends on its diameter and thereby its conductance $C_{kl}$. Larger conductance gives rise to larger maintenance power, hence the exponent $\gamma >0$. For real systems, the maintenance power is linear or sublinear \cite{HuCai2013}, implying that $\gamma \le 1$. The constant $c_0=(\tau_e)^2 /\gamma$ reflects the scale ratio of flow power and maintenance power. We use the scale $\tau_e=1$ so that $c_0=1/\gamma$. The prefactor $1/2$ compensates for double counting each edge in the sum, both as $(k,l)$ and $(l,k)$.

The gradient of the energy consumption \eqref{eq:Energy_general} has the components
\begin{align}
\frac{\partial E}{\partial C_{ij}} = - \frac{Q_{ij}^2}{C_{ij}^2} + \gamma c_0 C_{ij}^{\gamma-1},
\end{align}
where we use that $C_{ij}=C_{ji}$ and $Q_{ij}=-Q_{ji}$. Note also that terms containing inner derivatives $\partial Q_{kl} / \partial C_{ij}$ cancel due to mass balance \eqref{eq:massbalance}, see~\citet{Lu2022} for details.

A system minimizing energy consumption may adapt conductances according to a gradient descent rule
\begin{align}
\frac{\d }{\d t} C_{ij} = - \frac{\partial E}{\partial C_{ij}}~.
\end{align}
More generally, one may consider any adaptation rule
\begin{align} \label{eq:adaptation_general}
\frac{\d }{\d t} C_{ij} = - f(C_{ij}) \frac{\partial E}{\partial C_{ij}}
\end{align}
with arbitrary non-negative function $f$ which ensures $\d E / \d t \le 0$. Hu and Cai~\cite{HuCai2013} choose
$f(x) = c x^{2-\gamma}$ with a constant $c >0 $ resulting in
\begin{align}\label{eq:adaptationrule}
\frac{\d }{\d t} C_{ij} = c \left( \frac{Q_{ij}^2}{ C_{ij}^{\gamma+1}} - \gamma c_0 \right) C_{ij}.
\end{align}
By appropriate choice of units of time and conductance, we obtain values of the constants $c=1$ and $c_0= 1 / \gamma$. Then, since  $Q_{ij} = C_{ij} (p_i - p_j)$, we have
\begin{align}
\frac{\d }{\d t} C_{ij} = \frac{C_{ij}^2 (p_i-p_j)^2}{ C_{ij}^{\gamma}} - C_{ij}~.
\end{align}
Note that taking into account actual vessel lengths facilitates application of the model to real vascular networks \cite{KlemmMartens2023}; however, as a simplification, we assume all vessel segments (edges) of the network to have unit length. 

Extrema and saddles of the energy landscape are characterized by $ \nabla E = 0$, amounting to
\begin{align}
\frac{Q^2_{kl}}{C_{kl}} = C_{kl}^{\gamma}~.
\end{align}
using $\gamma c_0 = 1$.
The left hand side of this equation is the cost of dissipation caused by the direct flow between nodes $k$ and $l$; the term $C_{kl}^\gamma$ on the right hand side is the maintenance cost of the vessel segment. The inner equilibria of the dynamics, equivalent to the critical points of the energy, are thus characterized by equality of these two contributions to the energy consumption.

\subsection{Sink fluctuations}\label{sec:sink_fluctuations}

We consider systems with a single source node $r \in V$ having $h_r = 1$.
We denote the set of sink nodes with $S \subset V$ where $r \notin S$. The net flow at sink nodes is assumed to fluctuate in general. This is implemented by choosing one sink $s \in S$ and assigning $s$ a larger net flow (in absolute value) than the other sinks as
\begin{align}\label{eq:load_fluctuations}
  h_i = \begin{cases}
                       1 & \text{ if } i=r \\ 
  -\dfrac{1-\alpha}{|S|}-\alpha & \text{ if } i=s \\
  -\dfrac{1-\alpha}{|S|}        & \text{ if } i \in S\setminus\{s\}\\
  0                      & \text{ otherwise. }
\end{cases} 
\end{align}
The parameter $\alpha \ge 0$ determines the amplitude of fluctuations. For $\alpha=0$, fluctuations vanish with each sink $i \in S$ having constant outflow $h_i=-1/|S|$. The case where $\alpha =1$ corresponds to a single moving sink~\cite{Katifori2010} since all sinks except high-flow sink $s$ have zero net flow. When $\alpha>1$, sink nodes with $i \neq s$ are assigned $h_i >0$, so they act as sources; however, this is not a scenario we expect to occur in real systems. Yet, the following analysis is valid for all values $\alpha \ge 0$.

For stochastic sink fluctuations occurring on a time scale $T>0$, the net flows in all sinks are constant on all time intervals $I_n=[(n-1)T,nT[$ with $n \in \N$. Independently for each $n \in \N$, we draw the high-flow sink $s_n \in S$ uniformly at random (with probabilities $1/|S|$). 


\section{Analysis}

\rev{In this section, we discover non-local transitions where the ranking of solutions with respect to the energy functional $E$ changes, and thus switching between ground (minimum $E$) states occurs (see subsections E and F). The impact of these transitions in a system with finite time-scale fluctuations is studied in subsection G. We provide context for these findings by first reviewing the known saddle-node bifurcations giving rise to cyclic solutions \cite{kaiser2020discontinuous,KlemmMartens2023} (subsections A-C). For particular values of cost exponent $\gamma$, we reveal additional bifurcations (subsection D).}


\subsection{Triangular network motif with 1 source and 2 sinks}\label{sec:triangular_motif_model}

We consider a triangular network motif with one source $h_1=1$ and two fluctuating sinks $h_2$ and $h_3$. This motif has already been considered in previous studies~\cite{MartensKlemm2017,martens2019cyclic,KlemmMartens2023} as it serves as an elementary building block for larger networks in setups with one source and many sinks (terminal loads).
 
We assume that the drive $h_{2,3}f(t)$ fluctuates on a rapid time scale, i.e., $T=T(h_{2,3})\ll 1$.
The sources then obey $\langle h_2 \rangle-\langle h_3 \rangle  \rightarrow 0$ and no net pumping occurs between nodes $k=2$ and $k=3$. 

Accordingly, we seek solutions, $\langle C_{kl}\rangle$, averaged over rapid fluctuations with characteristic time scale $T$. We consider that $C_{ij}$ changes on a slow time scale, $C_{ij} \rightarrow \langle C_{ij}\rangle$ as $T \rightarrow 0$. From now on, we therefore use $\langle C_{kl}\rangle$ and $C_{kl}$ interchangeably and simplify notation by omitting $\langle\cdot\rangle$ around the conductances. Note that the symmetry of both the triangular network motif and of the (rapid) fluctuations imply that $\langle C_{12}\rangle = \langle C_{13} \rangle$. 

In this rapid fluctuation limit, the dynamics of the conductances is constrained to a two dimensional subspace with the dynamics governed by
\begin{subequations}\label{eq:triangular_motif_goveqn_symmetric}
\begin{align}
\label{eq:triangular_motif_goveqn_symmetric_a}
 \frac{\d }{\d t} C_{12} &=  C_{12}(C_{12}^{1-\gamma}\langle(p_1-p_2)^2\rangle - 1),\\
 \label{eq:triangular_motif_goveqn_symmetric_b}
 \frac{\d }{\d t} C_{23} &= C_{23}(C_{23}^{1-\gamma}\langle (p_2-p_3)^2 \rangle - 1).\
\end{align} 
\end{subequations}
The conductances are related via the mass balance~\eqref{eq:massbalance}, i.e.,
\begin{subequations}\label{eq:triangular_motif_massbalance}
\begin{align}
\label{eq:triangular_motif_massbalance_a}
 1&= C_{12}(p_1-p_2)+C_{13}(p_1-p_3),\\
\label{eq:triangular_motif_massbalance_b}
 h_2&= C_{12}(p_2-p_1)+C_{23}(p_2-p_3),\\
\label{eq:triangular_motif_massbalance_c} 
 h_3&= C_{13}(p_3-p_1)+C_{23}(p_3-p_2),\
\end{align}
\end{subequations}
where we may (without loss of generality) set the reference pressure $p_1=0$.
The mass balance is further simplified using the symmetry $C_{12}=C_{13}$  in Eqs.~\eqref{eq:triangular_motif_massbalance_a}-\eqref{eq:triangular_motif_massbalance_c}. We omit the details of the  analysis which was carried out in our previous studies~\cite{MartensKlemm2017,KlemmMartens2023} and instead summarize the main results in the two following Secs.~\ref{sec:triangular_motif_stationary} and \ref{sec:triangular_motif_bifurcutions_stability}.

Relations between the average squared terminal fluctuations, $h_1,h_2$ and $h_3$, and  the fluctuation amplitude $\alpha$ relevant to evaluating the pressure difference terms in Eq.~\eqref{eq:triangular_motif_goveqn_symmetric}, are derived in Appendix A.1. in Ref.~\cite{KlemmMartens2023} They are $\langle h_2^2 \rangle =  \frac{1+\alpha^2}{4}, \langle (h_2-h_3)^2 \rangle =  \alpha^2, \langle h_2 h_3 \rangle =  \frac{1-\alpha^2}{4}$.


\subsection{Stationary solutions of triangular motif}\label{sec:triangular_motif_stationary}

The triangular motif described by Eqs.~\eqref{eq:triangular_motif_goveqn_symmetric} has two types of stationary solutions as previously shown~\cite{KlemmMartens2023}. 
We do not explicitly calculate equilibria for \eqref{eq:triangular_motif_goveqn_symmetric}, and instead refer to our previous study\cite{KlemmMartens2023} for a detailed derivation; however, we note that the symmetry assumption $C_{12}=C_{13}$ in conjunction with imposing an  equilibrium condition to Eq.~\eqref{eq:triangular_motif_goveqn_symmetric} implies the key relations $\avg{(p_1-p_3)^2}=\avg{(p_1-p_2)^2}$.
These include the \emph{tree-like solution}, $\mathbf{B}_\wedge=(C_{12},C_{23})$, with
\begin{align}\label{eq:treelike_branch}
    \mathbf{B}_\wedge=\left(\langle h_2^2\rangle^{\frac{1}{1+\gamma}},0\right) = \left(\left(\frac{1+\alpha^2}{4}\right)^{\frac{1}{1+\gamma}},0\right),
\end{align}
which is always stable for $0<\gamma<1$; stable for $\alpha<\alpha_c$ with $\gamma=1$, and repellent for $1<\gamma\leq 2$;
and the \emph{cyclic solution}, $\mathbf{B}_\Delta=(C_{12},C_{23})$, implicitly defined via the following two relations valid for $\gamma\neq 1$,
\begin{subequations}\label{eq:cyclic_branch_gamma_neq_one}
\begin{align}
    \label{eq:cyclic_branch_gamma_neq_one_C23}
    C_{23} = C_{12}  (4- C_{12}^{-\gamma-1})^{\frac{1}{\gamma-1}} ~,\\
    \label{eq:cyclic_branch_gamma_neq_one_alpha}
    \alpha = (C_{12}+2C_{23}) C_{23}^\frac{\gamma-1}{2} ~.
\end{align}
\end{subequations}
For $\gamma=1$, one may instead find~\cite{MartensKlemm2017,KlemmMartens2023} the explicit solution
\begin{subequations}\label{eq:C23forgamma1finalequi}
\begin{align}\label{eq:C23forgamma1finalequi_a}
    C_{12} &= \frac{1}{\sqrt{3}},\\\label{eq:C23forgamma1finalequi_b}
    C_{23} &= \frac{1}{2} \left(\alpha-\sqrt{\frac{1}{3}}\right).\
\end{align}
\end{subequations}
These relations produce the bifurcation curves shown in Fig.~\ref{fig:figure2}. For $\gamma\neq 1$, one may treat the variable $C_{12}\geq 0$ as a free parameter and use \eqref{eq:cyclic_branch_gamma_neq_one_C23} and \eqref{eq:cyclic_branch_gamma_neq_one_alpha} generate the tuples $(\alpha(C_{12}), C_{12})$ and $(\alpha(C_{12}),C_{23}(C_{12}))$.


\subsection{Bifurcations and stability diagram}\label{sec:triangular_motif_bifurcutions_stability}
\begin{figure}[htp!]
 \centering
 \includegraphics[width=\columnwidth]{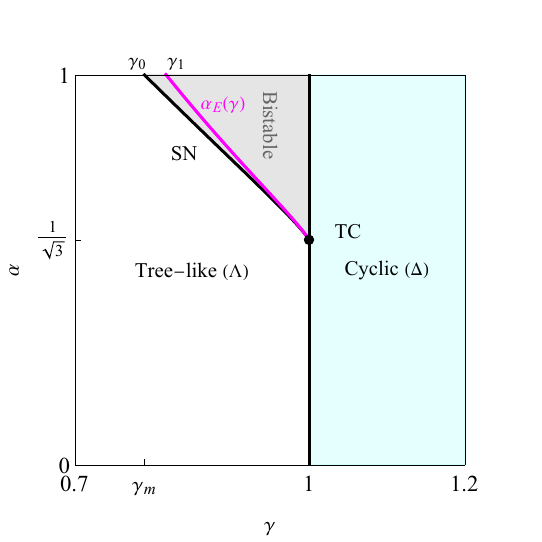}
 \caption{\label{fig:figure1}Stability diagram for the triangular network motif in the regime of rapid fluctuations ($T \ll 1$). The curves of the saddle-bifurcation (SN) and of $\alpha_E(\gamma)$  intersect $\alpha=1$ when $\gamma_0\approx 0.785$ and $\gamma_1\approx 0.816$, respectively. Note that the diagram extends to $\alpha >1$ continuously, but this regime is ignored for physical reasons.
}
\end{figure}

We summarize  the results on the bifurcation behavior and stability analysis for stationary solutions occurring in the triangular network motif~\cite{KlemmMartens2023,MartensKlemm2017}.
Bifurcation diagrams and phase diagrams for selected values of $\gamma$ are shown in Fig.~\ref{fig:figure1} and in Fig.~\ref{fig:phaseportraits} panels a) to c), respectively. 

\begin{figure*}[htp!]
 \centering
 \begin{overpic}[width=0.33\textwidth]{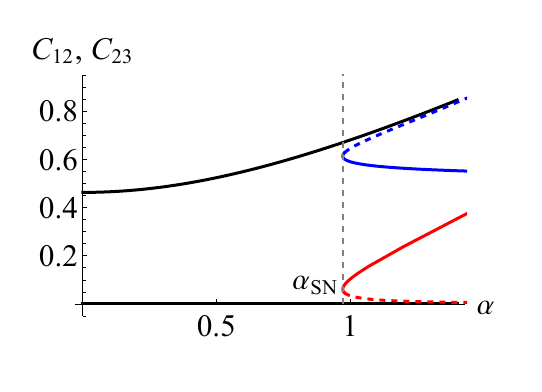}
 \end{overpic}
 \begin{overpic}[width=0.33\textwidth]{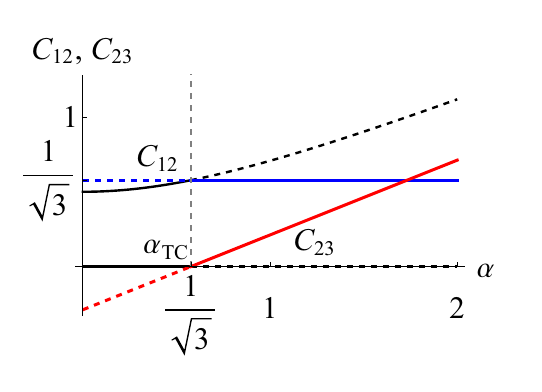}
 \end{overpic}
 \begin{overpic}[width=0.33\textwidth]{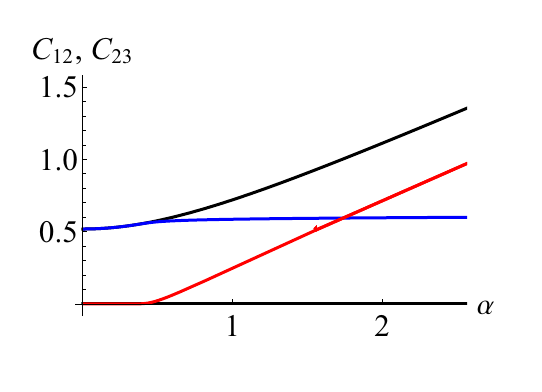}
 \end{overpic}
 \\
 \begin{overpic}[width=0.33\textwidth]{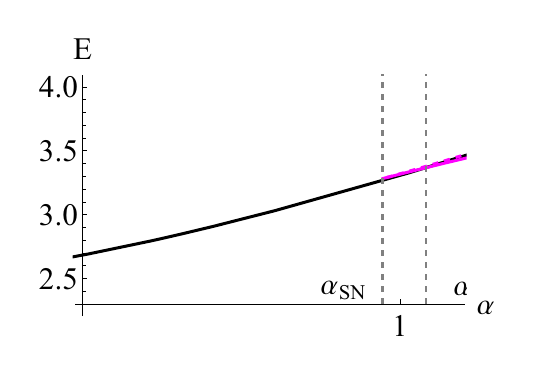}
    \put(17,47){\includegraphics[width=0.16\textwidth]{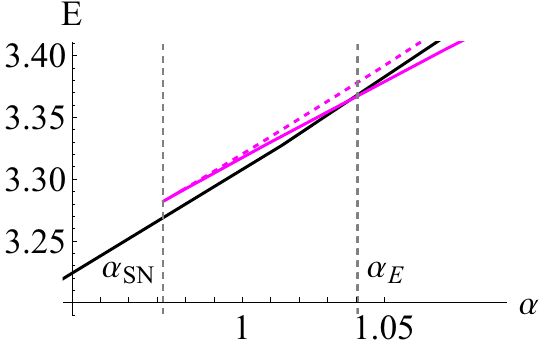}}
 \end{overpic}
 \begin{overpic}[width=0.33\textwidth]{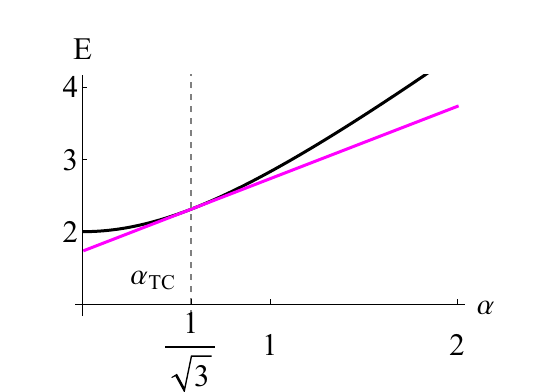}
 \end{overpic}
 \begin{overpic}[width=0.33\textwidth]{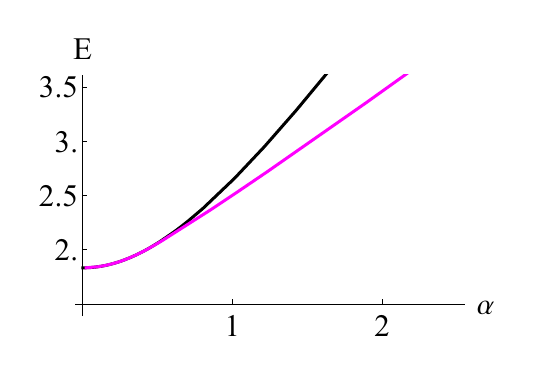}
 \end{overpic}
 
 \caption{\label{fig:figure2}Bifurcation diagrams (top tow) and energy diagrams (bottom row) for the triangular network motif with $T\ll 1$. Diagrams are shown for $\gamma=0.8$ (left) $\gamma=1$ (middle) and $\gamma=1.11$ (right). Tree-like solution branches and associated energy are shown as black curves. Cyclic solutions for are shown in blue ($C_{12}$) and red ($C_{23}$), and associated energy levels in magenta. Solution branches / energies associated with the stable and unstable (saddle) branches are shown as solid and dashed curves, respectively.\\
 }
\end{figure*}

\begin{figure*}[htp!]
 \begin{overpic}[width=0.32\textwidth]{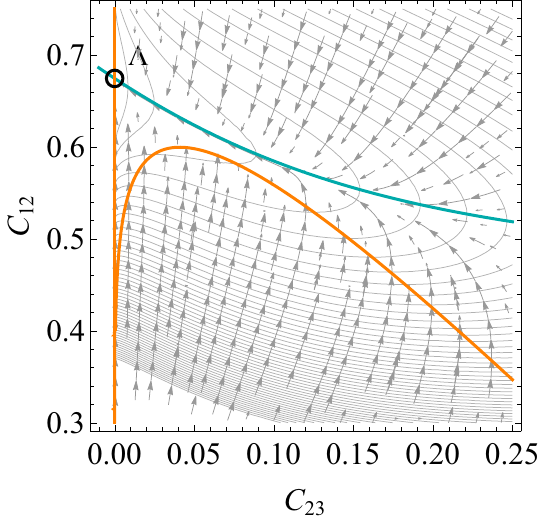}
  \put(70,90){a) $\gamma=0.76$}
 \end{overpic}
 \begin{overpic}[width=0.32\textwidth]{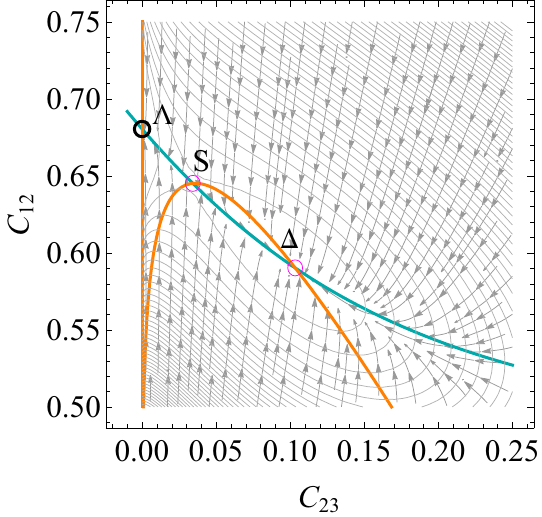}
 \put(70,90){b) $\gamma=0.8$}
 \end{overpic}
 \begin{overpic}[width=0.32\textwidth]{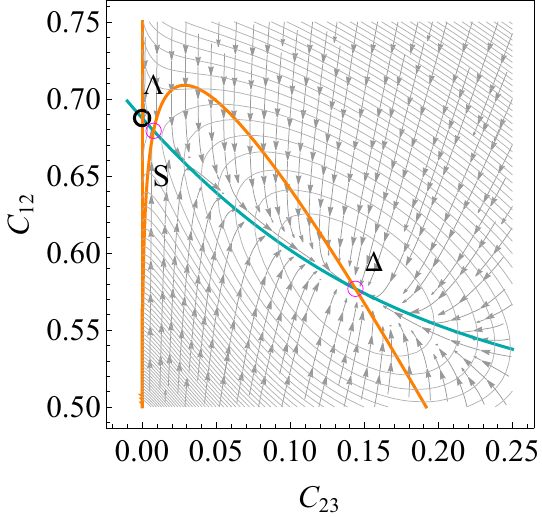}
 \put(70,90){c) $\gamma=0.85$}
 \end{overpic}

 \medskip

 \begin{overpic}[width=0.32\textwidth]{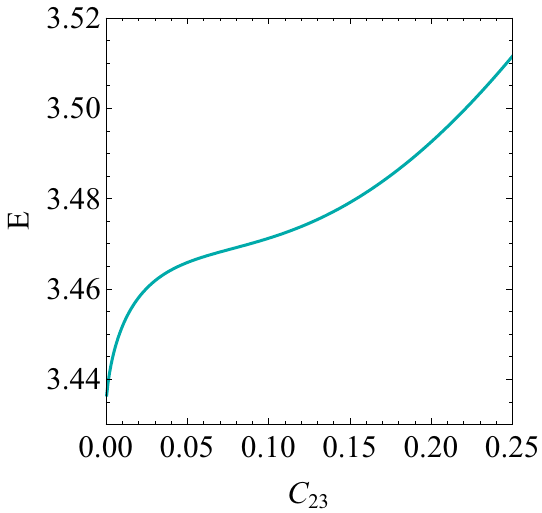}
  \put(68,85){d) $\gamma=0.76$}
 \end{overpic}
 \begin{overpic}[width=0.32\textwidth]{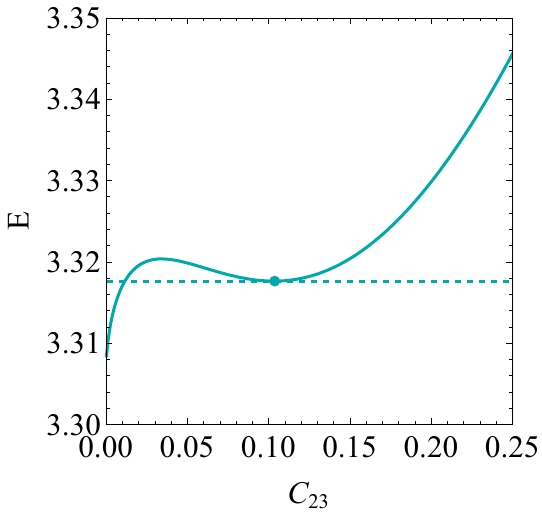}
 \put(68,85){e) $\gamma=0.8$}
 \end{overpic}
 \begin{overpic}[width=0.32\textwidth]{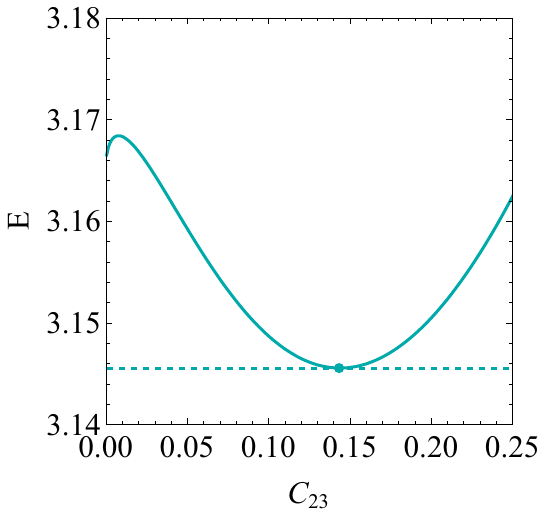}
 \put(68,85){f) $\gamma=0.85$}
 \end{overpic}

 \caption{\label{fig:phaseportraits}  Phase portraits for the triangular network motif with ($T\ll 1$) are shown in the top row (panels a)-c)), with parameter values $\alpha=1$ and $\gamma$ specified in the panels. $C_{12}$-nullclines  and $C_{23}$ nullclines are shown as blue and orange curves, respectively. Tree-like ($\Lambda$) and cyclic ($\Delta$) equilibria are highlighted as black and magenta circles, respectively. The lower row (panels d)-f)) show the energy consumption $E$ along the $C_{12}$-nullcline as a function of $C_{23}$. The dotted horizontal line indicates the value of $E$ at the local minimum corresponding to the stable cyclic equilibrium solution. 
}
\end{figure*}

For $\gamma < 1 $, the tree-like  solution $B_\Lambda $ \eqref{eq:treelike_branch} is always present, and stable for any value of $\alpha$. The cyclic solution emerges in a saddle-node bifurcation (SN) at $\alpha=\alphaSN(\gamma)$.
For $\alpha<\alphaSN$, there is only a stable tree-like solution $B_\Lambda $, see Fig.~\ref{fig:phaseportraits} panel a) and Fig.~\ref{fig:figure2} panel a); 
for $\alpha>\alphaSN$, we see a stable node ($\Delta$) and a saddle (S) corresponding to cyclic solutions,and the  tree-like solution ($\Lambda$), see Fig.~\ref{fig:phaseportraits} panel b) and Fig.~\ref{fig:figure2} panels b) or c). As a consequence, tree-like and one cyclic solution stably co-exist above the bifurcation with $\alpha>\alphaSN$.
An approximate condition for the saddle-node bifurcation was derived previously in Ref.~\citenum{KlemmMartens2023}, i.e., the value for $C_{12}$ at the saddle node bifurcation is well approximated by \footnote{The result of Eq.~\eqref{eq:C12_SN_approx} was rendered incorrectly as Equation~(37) in Reference~\citenum{KlemmMartens2023}. An erratum has been submitted.}
\begin{align} \label{eq:C12_SN_approx}
    C_{12}^\text{\, SN} = \left(4-\left(\frac{2(1-\gamma)}{3(1+\gamma)}\right)^{\gamma-1} \right)^{-1/(1+\gamma)}~.
\end{align}
Substitution of this value into Eqs.~\eqref{eq:cyclic_branch_gamma_neq_one_C23} and \eqref{eq:cyclic_branch_gamma_neq_one_alpha} yields the conductance $C_{23}^\text{\,SN}$ and the critical value $\alpha=\alphaSN$ at the SN bifurcation in the parameter regime where $\gamma<1$. 

For $\gamma = 1$, the tree-like and cyclic solution branches collide in a transcritical bifurcation (TC). This bifurcation point occurs where the solution changes sign, and \eqref{eq:C23forgamma1finalequi_b} determines a critical parameter value, 
\begin{align}
    \alphaTC = \sqrt{1/3}~,
\end{align}
in agreement with previous studies~\cite{MartensKlemm2017,martens2019cyclic}. Thus, below the trans-critical threshold, the only stable solution is tree-like; above the threshold, the only stable solution is cyclic.

Finally, for $\gamma>1$, the saddle-node bifurcation is annihilated in the transcritical bifurcation point TC and the tree-like and cyclic solutions stably co-exist for any value of $\alpha$, see Fig.~\ref{fig:figure2} panel c).  Note that the tree-like and cyclic solutions display nearly identical values for $C_{23}$ below a specific value of $\alpha$ and only begins to substantially differ above a certain $\alpha$-value. 

The stability diagram from these bifurcations is shown in Fig.~\ref{fig:figure1}. We remind the reader that not all values of $\alpha$ and $\gamma$ are of interest to us, due to certain limitations in their physical interpretations. First, we limit our attention to $\alpha \leq 1$ to avoid sign reversal in the terminal fluctuations (see Sec.~\ref{sec:sink_fluctuations}).
Second, for the exponent value $\gamma >1$, the maintenance power of a vessel segment scales superlinearly with its conductance. This case appears less realistic than the linear ($\gamma=1)$ and sublinear ($\gamma<1$) ones. Under $\gamma>1$, maintenance power may always be lowered by replacing a vessel segment with several parallel segments while keeping total conductance. 

\subsection{Additional saddle-node bifurcations}
For specific values of $\gamma$, the governing equations allow for additional saddle-node bifurcations. To see this, consider Eq.~\eqref{eq:C12_SN_approx},
which is of the general form
$C_{12}=(-x)^\beta = (-1)^\beta |x|^\beta = e^{i\pi \beta} |x|^\beta$,
with $x=(2/3(1-\gamma)/(1+\gamma))^{\gamma-1}-4$ and $\beta=-1/(1+\gamma)$. Exponentiation with values $\beta \notin \Z$ will result in a complex solution which is to be rejected. However, exponents with $\beta = k$ where $ k  \in \Z$ result in a real-valued $C_{12}$. Furthermore, $C_{12}>0$ if $\beta=2k$; but $C_{12}<0$ if $\beta=2k+1$. Thus, we expect additional saddle-node bifurcations producing physically realistic values exactly when
\begin{align}
 \beta = \frac{-1}{1+\gamma}=2k,\quad k \in \Z.
\end{align}
Note that almost the same condition (apart from a sign flip)  for the exponent of Eq.~\ref{eq:cyclic_branch_gamma_neq_one_C23} controls whether $C_{23}$ is complex- or real-valued.
These additional saddle-node bifurcations are destroyed upon perturbation in $\gamma$ and are thus not considered to be structurally robust. Nevertheless, we mention them here as they may avoid confusion for future investigators.

\subsection{Energy consumption for the triangular motif}

We now turn our attention to the model in terms of minimizing energy consumption $E$ as given by \eqref{eq:Energy_general}. Relative values for $E$ evaluated at different equilibria allow for a more global characterization of stability, which is particularly useful to discuss the system behavior in the presence of multistability (occurring for $\alpha>\alphaSN$ and $\gamma\leq 1$) and of noise-driven dynamics in the regime of non-rapid slow fluctuations, which we will discuss later in Sec.~\ref{sec:sink_fluctuations}.

Let us obtain an explicit expression for the energy consumption $E$ of the triangular network motif. Using $Q_{ij}=C_{ij}(p_i-p_j)$, we can rewrite \eqref{eq:Energy_general} in terms of pressure differences. Additionally, the symmetry of load fluctuations between nodes 2 and 3 implies $C_{13}=C_{12}$, and so the energy consumption becomes
\begin{align} \label{eq:Et_with_p}
\begin{split}
 E &= 2 C_{12} \avg{(p_1-p_2)^2} + C_{23} \avg{(p_2-p_3)^2} \\
 &+ \frac{2}{\gamma} C_{12}^\gamma + \frac{1}{\gamma}C_{23}^\gamma~,
 \end{split}
\end{align}
where we also replaced $c_0 = \gamma^{-1}$. The expectation values of quadratic pressure differences have been obtained using mass balance, see Eqs.~(17) and (18) in Reference~\citenum{KlemmMartens2023}, which are 
\begin{align}\label{eq:pl_ph}
  \avg{(p_2-p_3)^2} = \frac{\alpha^2}{(C_{12} + 2 C_{23})^2} =: X
\end{align}
and
\begin{align}\label{eq:p2_p1}
    \avg{ (p_2 - p_1)^2} = \frac{1}{4} \left(\frac{1}{C_{12}^2} + X\right)~.
\end{align}
Inserting the former expressions into \eqref{eq:Et_with_p}, we have
\begin{align}\label{eq:Et1}
 E &= \frac{1}{2 C_{12}} + \frac{\alpha^2}{2(C_{12}+2C_{23})} + \frac{2}{\gamma} C_{12}^\gamma + \frac{1}{\gamma}C_{23}^\gamma~.
\end{align}

We evaluate $E$ along the $C_{12}$-nullclines (blue curves in Fig.\ref{fig:phaseportraits} panels (a)-(c)), as shown in Figure~\ref{fig:phaseportraits} in panels (d)-(f). In particular, maximum and minima of $E$ give the energy consumption at the saddle and stable fixed points. Fixing $\alpha=1$ and choosing subcritical $\gamma=0.76 < \gamma_0$, the unique minimum of $E$ is at the tree-like solution, see Figure~\ref{fig:phaseportraits}(d). For values $\gamma=0.80 > \gamma_0$ and $\gamma=0.85 > \gamma_0$, i.e.\ above the saddle-node bifurcation, the stable cyclic solution is a local minimum; the unstable cyclic solution (saddle) is a local maximum of $E$ as seen in panels (e) and (f) of Figure~\ref{fig:phaseportraits}. A change of behaviour occurs between $\gamma=0.80$ and $\gamma=0.85$ at $\gamma=\gamma_1$. For $\gamma<\gamma_1$, the tree-like solution with $C_{23}=0$ is the unique global minimum of $E$. For $\gamma>\gamma_1$, the stable cyclic solution uniquely globally minimizes $E$. Thus in addition to the saddle-node bifurcation at $\gamma=\gamma_0$, the system undergoes another transition at $\gamma_1$, where the ground state changes from the tree-like fixed point to the stable cyclic fixed point. 

The phenomenon of switching ground state generalizes to parameter values $\alpha \le 1$. In the stability diagram in Figure~\ref{fig:figure1}, the curve $\alpha_E(\gamma)$ further subdivides the parameter regime of bistability. Above the curve, i.e.\ for $\alpha>\alpha_E(\gamma)$, the stable cyclic solution is the ground state configuration. Below the curve, the tree-like solution globally minimizes $E$.


\subsection{System with a multitude of local minima of energy}

\begin{figure*}[htp!]
\includegraphics[width=0.8\textwidth]{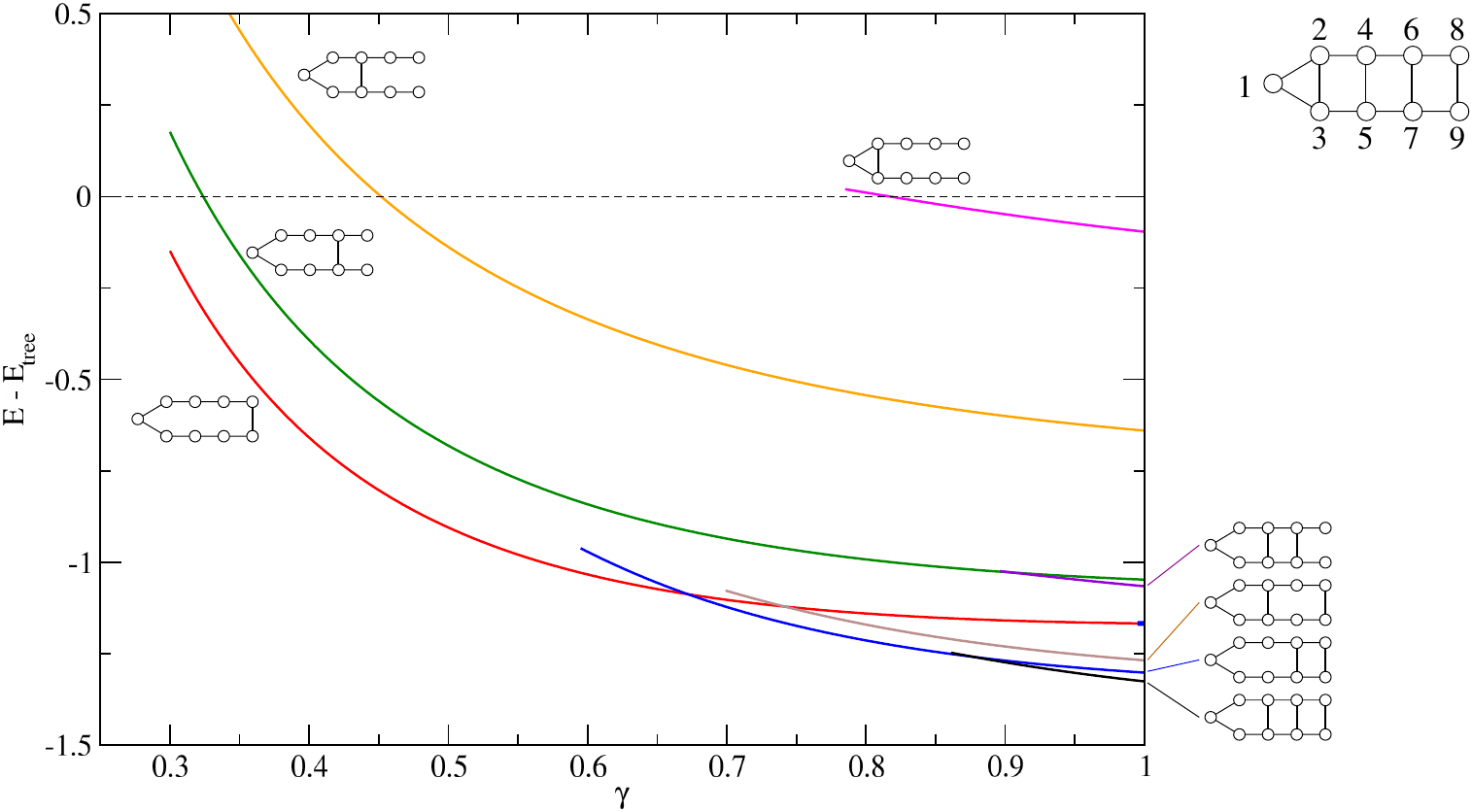}
\caption{\label{fig:gammanullclt_0}
Energies of stable equilibria in dependence of the cost exponent $\gamma$ in a network with 1 source (node 1) and 8 sinks, shown in the upper right corner. Fluctuation amplitude is set to $\alpha=1.0$. For each value of $\gamma$ where a stable cyclic equilibrium exists, we plot the energy difference $E-E_\text{tree}$ between cyclic solution and tree solution. The tree solution (dashed horizontal line) is the one without any of the cross-edges $\{2,3\}$, $\{4,5\}$, $\{6,7\}$, $\{8,9\}$.
}
\end{figure*}

Let us consider a larger network with more cycles and thus more possibilities for the presence or absence of cross-edges. Figure \ref{fig:gammanullclt_0} shows the test network with 9 nodes and 12 edges, where we denote the four vertically drawn edges as cross-edges, the remaining ones as tree-edges. There are $2^4 = 16$ combinations of conducting / non-conducting cross-edges; 15 of these are cyclic (with at least one cross-edge conducting); the remaining combination is the tree without any conducting cross-edges. 

Out of the 15 cyclic combinations, 7 are not observed as stable equilibria for any values of $\gamma \in [0,1]$, while $\alpha=1.0$ remains constant. For each of the remaining 8 cyclic combinations, Figure \ref{fig:gammanullclt_0} displays the energy values for the stable equilibrium on the $\gamma$-interval where the cyclic equilibrium exists. As the parameter $\gamma$ is varied, the ranking of equilibria by energy varies. In particular, which equilibrium constitutes the ground state, is strongly parameter-dependent. At $\gamma=1.0$, the ground state is the one with all cross-edges except $\{2,3\}$. At $\gamma \approx 0.86$, $\gamma \approx 0.67$, the ground state configuration switches between different cyclic equilibria. Below $\gamma \approx 0.30$, the ground state is the tree.


\subsection{Sink nodes fluctuating on finite time scale}

\begin{figure*}[htp!]
\includegraphics[width=0.8\textwidth]{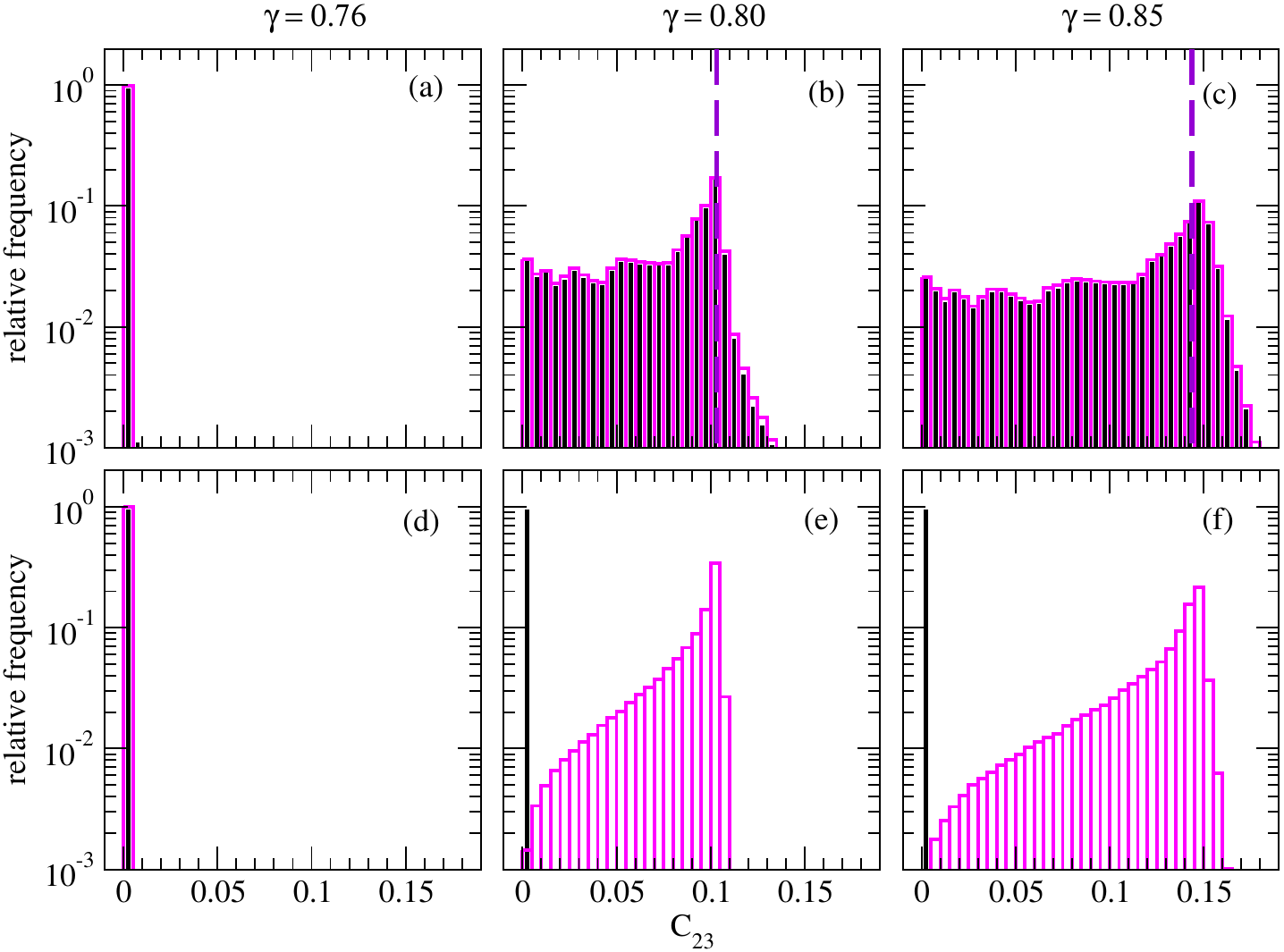}
\caption{\label{fig:gfluci_0}
Histograms of values of $C_{23}$ under load fluctuations at sink nodes $2$ and $3$ with $\alpha=1.0$ (single moving sink) and $\gamma \in \{0.76, 0.80, 0.85\}$ for the system with 1 source and 2 sinks. The upper row of panels (a-c) have fluctuation time scale $T=1.0$, the lower row of panels \rev{(d-f)} have $T=0.5$. In each panel, histograms are drawn for the dynamics with initial condition $(C_{12},C_{13},C_{23})=(1,1,1)$ (open magenta bars), and for initial condition $C_{12}=C_{13}={0.5}^{1/(1+\gamma)}$ and $C_{23}={10}^{-6}$ (solid black bars). The latter initial condition corresponds to the tree-like solution \eqref{eq:treelike_branch} perturbed by adding ${10}^{-6}$ to the $C_{23}$ component. Dashed vertical lines indicate the $C_{23}$ value in the stable cyclic solution, see also Fig.~\ref{fig:phaseportraits} panels b) and c). Euler integration is performed with a time step $\Delta t = 10^{-4}$ until time $t$ reaches $10^5$. We have verified that histograms remain unchanged (up to sampling error) when varying random number sequences for the sink fluctuations.}
\end{figure*}

\begin{figure}[htp!]
\begin{overpic}[width=0.49\textwidth]{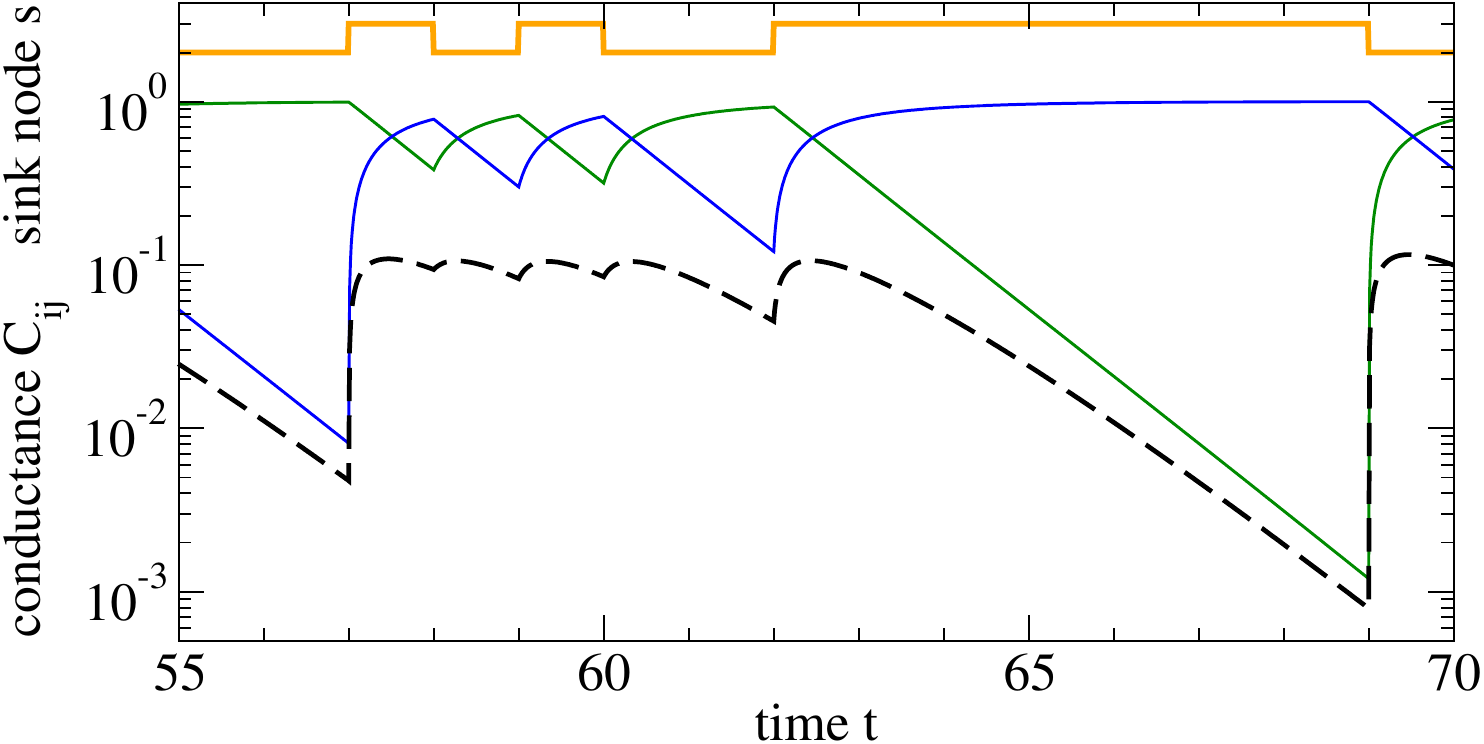}
 \put(7,46){\tiny $s=2$}
 \put(7,48){\tiny $s=3$}
\end{overpic}

\caption{\label{fig:gfluc_0}
Example of conductances' time evolution under slow fluctuations in the 3-node system with source node 1 and sink nodes $\{2,3\}$.  The thick curve indicates which one of the two sinks $s \in \{2,3\}$ is active at any given time $t$. The evolution of conductances $C_{23}$ (dashed black), $C_{12}$ (solid green) and $C_{13}$ (solid blue) are show.
Parameter values are $\gamma=0.8$ and $\alpha=1.0$.
}
\end{figure}

We now relax the time scale separation between the sink fluctuations and the network adaptation. Accordingly, we no longer require that $T\ll 1$, and  we analyze the tree-node motif under stochastic fluctuations with time scale $T$ as described in Sec.~\ref{sec:sink_fluctuations}. We choose the fluctuation amplitude $\alpha=1.0$. Thus at any point in time, we either have $h_2(t) = -1$ and $h_3(t)=0$, or $h_2(t)=0$ and $h_3(t)=-1$ as the net flow at sink nodes 2 and 3. We integrate the dynamics for conductances $(C_{12},C_{13},C_{23})$ according to Eq.~\eqref{eq:adaptationrule}. We record the distribution of $C_{23}$ values and plot histograms in Fig.~\ref{fig:gfluci_0}.

For fluctuation time scale $T=1.0$ (upper row of panels in Fig.~\ref{fig:gfluci_0}), the same distributions are obtained regardless of chosen initial conditions. At $\gamma=0.76$, in the subcritical regime with respect to the saddle-node bifurcation, the configurations encountered are tree-like with $C_{23}$ close to zero. For the supercritical choices $\gamma=0.80$ and $\gamma=0.85$, mostly cyclic configurations are obtained, see Fig.~\ref{fig:gfluci_0})(b,c). In these panels, for comparison we demark the value of $C_{23}$ corresponding to the stable cyclic equilibrium obtained for the rapid fluctuation limit ($T\ll 1$) with a vertical dashed line. The distribution of $C_{23}$ peaks remain (relatively) close to this value for both $\gamma=0.80$ and $\gamma=0.85$. For $T=1.0$ and all values of $\gamma$, the distribution of $C_{23}$ is independent of the initial condition.

At shorter fluctuation time scale $T=0.5$ and supercritical $\gamma \in \{0.80,0.85\}$, the observed $C_{23}$ distribution does depend on the initial condition, see Fig.~\ref{fig:gfluci_0} panels (e) and (f). Using the initial conditions close to the tree-like solution, we observe only tree-like configurations with values of $C_{23}$ concentrating towards values close to zero. As an initial condition sufficiently far from the tree-like configuration we use $(C_{12},C_{13},C_{23})=(1,1,1)$ and observe mostly tree-like configurations.

Figure \ref{fig:gfluc_0} shows an example of the conductances evolving under slow sink fluctuations. Notice that when sink $s$ is active during a given time interval, the direct conductance $C_{1s}$ from source node 1 to active sink $s$ is strengthening, while the other two conductances are exponentially decreasing. In Fig.~\ref{fig:gfluc_0}, this is seen for time interval $[62.0, 69.0]$ where sink $s=3$ is constantly active. After switching to sink $s=2$, both $C_{12}$ and $C_{23}$ increase rapidly. The increase of $C_{23}$ is due to large value of $C_{13}$ causing a low pressure difference between source 1 and sink 3 at this time. This makes the direct connection via $C_{12}$ and the indirect connection via sink 3 involving $C_{23}$ almost equally cost-efficient in establishing flow from source 1 to current sink 2. The mechanism explains the broad fluctuations in $C_{23}$ at $T\gg 1$. The cross-conductance intermittently assumes values close to zero without being drawn into the equilibrium at $C_{23}=0$.

Note that considerations using the energy function are only (strictly) valid for the deterministic limit (or $T\ll 1$); in the case of finite time fluctuations, the energy function is no longer a Lyapunov function for the system.


\section{Conclusions and Discussion}

\begin{figure*}[htp!]
\includegraphics[width=0.8\textwidth]{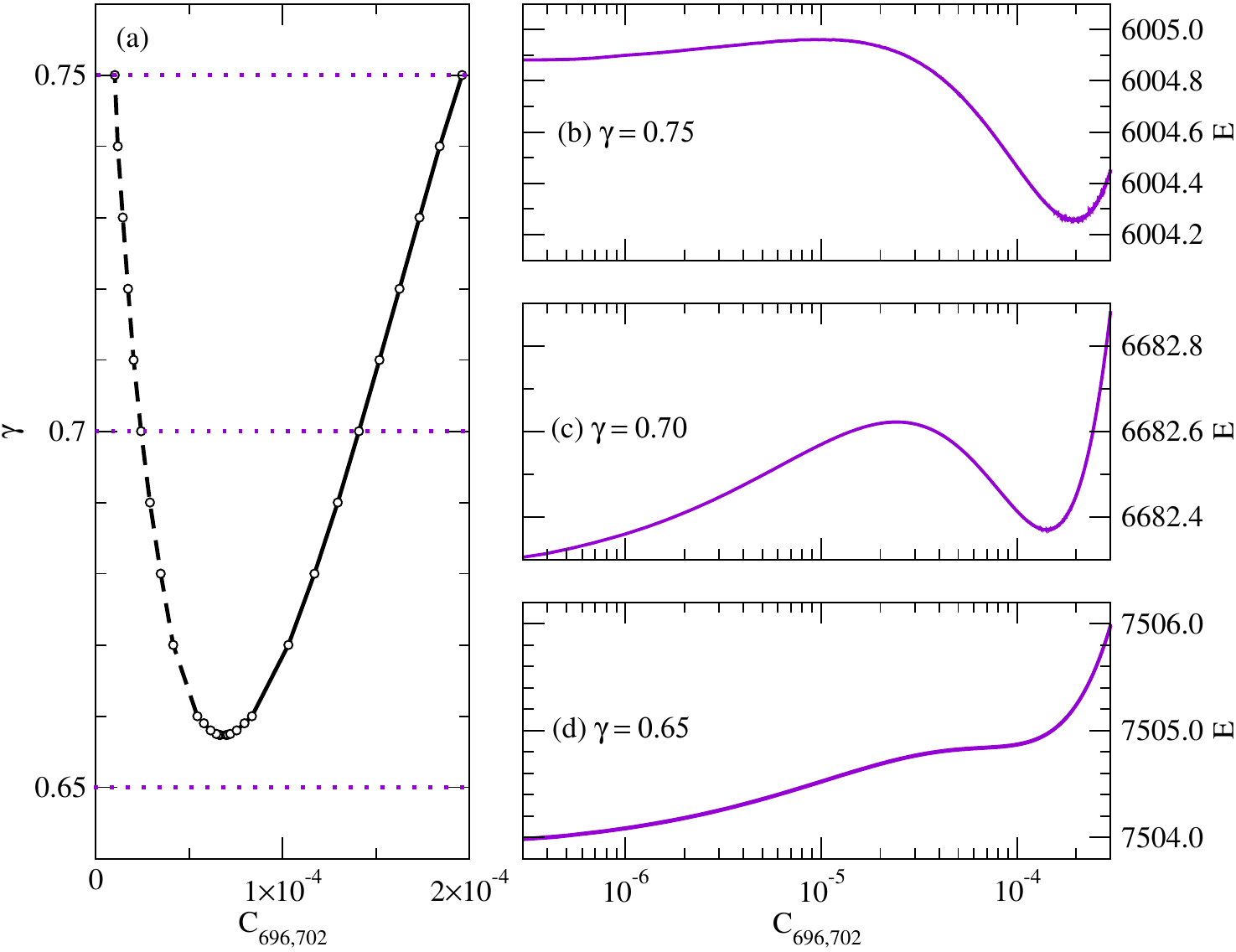}
\caption{\label{fig:gammanullclb_0}
Saddle-node bifurcation and energy landscape in an empirical vascular network. (a) Bifurcation diagram for cyclic ($C_{696,702}>0$) fixed point solutions in dependence of cost exponent $\gamma$. 
(b,c,d) Energy plotted along curves through configuration space. Each curve contains the stable fixed point and the saddle-point seen in panel (a), showing as a local minimum / maximum respectively. For each $\gamma \in \{0.75,0.70,0.65\}$, the points on the curve fulfill $\d C_{kl} / \d t =0$ for all $\{k,l\} \neq \{696,702\}$, whereas $C_{696,702}$ chosen in the range $[3 \times 10^{-7}, 3 \times 10^{-3}]$ is used to parameterize the curve. The network is from the data set by Blinder {\it et al.}\ \cite{Blinder2010} labeled 012208 with 826 nodes and 855 edges. Fluctuations have $\alpha=1.0$, and are to be understood in the rapid limit, $T\ll 1$. In this setting, the stable solutions with positive conductivity on the vessel between nodes 696 and 702 are the first ones to disappear by a saddle-node bifurcation as $\gamma$ is lowered from 1.0 towards zero, see also the preliminary work in \citenum{KlemmMartens2023}.
}
\end{figure*}

The model by Hu and Cai describes a vascular network with adaptive conductances at two levels, namely (i) an energy landscape established on the space of conductance values, with the energy function $E$ being the maintenance cost plus the power needed to sustain the pipe flow, summed over all vessel segments; and (ii) a system of coupled differential equations describing the time evolution of the conductances. Levels (i) and (ii) are linked through the fact that the conductances' evolution is chosen such as to never increase the value of the energy function $E$. Thus, each critical point of $E$ is a fixed point of the dynamics. In particular, each local minimum of $E$ is a stable fixed point (Fig.~\ref{fig:phaseportraits}).

The model shows multistability already for the system of three nodes with one source and two sink nodes. The energy function $E$ is a means of discerning the stable solutions. This establishes another critical curve in parameter space where a \rev{switch} in ground state occurs: the two stable solutions, tree-like and cyclic, exchange their ranking in terms of energy and thus their thermodynamic stability. This phenomenon is illustrated in Fig.~\ref{fig:gammanullclt_0} for an example network with 1 source and 8 sinks.

The presence of multistability brings up the question, which solution is approached by the dynamics. Specifically, is the energetically lower solution always preferred? We made this question precise by studying the effect of stochastic sink fluctuations, occurring on a time scale of order $T$ compared to the time scale of the conductance dynamics which is of order $1$. These fluctuations were averaged over when they were assumed to occur on a faster time scale compared to the dynamics of the conductances (i.e., $T\ll 1$), resulting in the deterministic dynamics given by Eqs.~\eqref{eq:triangular_motif_goveqn_symmetric} and \eqref{eq:triangular_motif_massbalance} (see results in Figs.~\ref{fig:figure1}-\ref{fig:phaseportraits} and \ref{fig:gammanullclb_0}). However, when the associated time scale separation is relaxed, stochasticity is introduced in the dynamics (Fig.~\ref{fig:gfluci_0}). The main result of the analysis is that the system exhibits mostly cyclic configurations even when the tree-like solution is energetically favourable in the deterministic limit for sufficiently large fluctuation strength, $\alpha$.

A question related to multistability is thus how large the basin of attraction is for the tree-like and cyclic solutions, respectively. 
The likelihood that the vascular network remains in the most energetically preferable state depends on the state's stability against significant perturbations. While linear stability analysis gives local information of stability, basin stability provides a more global view on this issue~\cite{menck2013basin}. It is not straight forward to define a trapping region (such as separatrices), especially while incorporating varying values for the system parameters. Therefore, we did not attempt the to devise a measure for the basin volume; however, inspection of the phase portraits in Fig.~\ref{fig:phaseportraits} panels b) and c) displays that the phase space is subdivided into regions of attraction by the separatrices emanating from the saddle point $S$, i.e., the stable and unstable manifolds. Thus, we may estimate that basin size~\cite{martens2016basins} for the tree-like solution scales with the distance between the stable fixed point $\Lambda$ and $S$; and that basin size for the cyclic scales like the distance between $S$ and the stable node $\Delta$. Note that all three fixed points more or less lie on a straight line. Thus, qualitatively we can see that the basin size for cycles $\Delta$ is comparatively much larger than the one for $\Lambda$, which at least for the shown parameters explains the higher likelihood to reside near the cyclic attractor when terminal fluctuations are sufficiently slow.

Shifting focus to larger systems, we have also analyzed the energy landscapes and dynamics on a large system based the vascular networks empirically obtained by Blinder {\it et al.}~\cite{Blinder2010}. Previous analysis by the authors~\cite{KlemmMartens2023}, considering the rapid fluctuation limit $T\ll 1$, had shown that a saddle-node bifurcation creates a pair of fixed point solutions with non-zero conductance for each cross-edge in a large network, analogous to the case of the triangular motif. 
Figure~\ref{fig:gammanullclb_0}(a) shows an example of such a bifurcation in an empirical network~\cite{Blinder2010} for $\alpha=1$ with $T\ll 1$. Similar to the energy function of the small system evaluated along the nullclines in Figure~\ref{fig:phaseportraits}, panels (b), (c), and (d) of Figure~\ref{fig:gammanullclb_0} show slices through the energy landscape for the large network. The main observation from the three-node motif is reproduced in the large network: Close to the bifurcation, the tree-like solution lies energetically lower than the cyclic solution; however, this ranking switches when tuning the parameter $\gamma$ farther away from the transition. We conjecture that the saddle-node bifurcations and the additional change of behaviour  in terms of the energy minimality of solutions are generic features of the model on all networks involving cycles.

For future research, it could be interesting to also consider other types of time-dependent networks. In particular, how are energy configurations altered in co-evolving networks where edges are dynamically created/deleted~\cite{Gross2008,berner2023adaptive}, such as in growing supply networks~\cite{ronellenfitsch2016global}?

\begin{acknowledgments}
K.K. acknowledges support from Project No. PID2021-122256NB-C22 funded by
MCIN/AEI/10.13039/501100011033/FEDER, UE.
We gratefully acknowledge financial support from the Royal Swedish Physiographic Society of Lund.
\end{acknowledgments}

\section*{Data Availability Statement}

Data sharing is not applicable to this article as no new data were created or analyzed in this study.


\bibliography{artnrg}
\end{document}